\documentclass[conference,a4paper]{IEEEtran}
\usepackage{balance}
\usepackage{import} 

\usepackage[utf8]{luainputenc}

\usepackage[T1]{fontenc}

\usepackage{graphicx}
\usepackage[dvipsnames]{xcolor}

\usepackage[english]{babel}
\addto\captionsenglish{\renewcommand{\figurename}{Fig.}}
\addto\captionsenglish{\renewcommand{\tablename}{Tab.}}
\usepackage{csquotes}

\usepackage{newtxtext}
\usepackage{amsthm}
\usepackage[slantedGreek]{newtxmath}
\usepackage[OMLmathsfit]{isomath}
\DeclareMathAlphabet{\mathbfsf}{\encodingdefault}{\sfdefault}{bx}{n}
\usepackage{bm}
\usepackage{envmath}
\usepackage{mathtools}
\usepackage{commath}

\usepackage[caption=false,font=footnotesize]{subfig}

\usepackage{booktabs}
\usepackage{footmisc}  

\usepackage{url}

\theoremstyle{definition}

\theoremstyle{plain}

\theoremstyle{remark}

\usepackage{lineno}
\modulolinenumbers[5]
\usepackage{todonotes}
\usepackage{umoline}

\usepackage{pgfplots}
\usepackage{pgfplotstable}
\pgfplotsset{compat=newest}
\pgfplotsset{plot coordinates/math parser=false}
\newlength\figureheight
\newlength\figurewidth
\pgfplotsset{every axis plot/.append style={line width=1.5pt},
    legend style={font=\footnotesize, 
        text height=1.0ex,
        draw=black,
        fill=white,
        legend cell align=left}}

\newcounter{subequation}
\newlength\mtabskip\mtabskip=-1.25cm

\def\mtabLong{long}

\newcommand{\mr}{\mathrm}

\newcommand{\veg}[1]{\bm{#1}}     
\newcommand{\mat}[1]{\mathsfbfit{#1}} 
\renewcommand{\vec}[1]{\mathsfbfit{#1}} 
\newcommand{\vecop}[1]{\bm{\mathcal{#1}}} 

\newcommand{\jm}{\mathrm{j}}  

\newcommand{\e}{\mathrm{e}}

\newcommand\restr[2]{{
        \left.\kern-\nulldelimiterspace 
        #1 
        \vphantom{|} 
        \right|_{#2} 
}}

\newcommand\rst[3]{{
        \left.\kern-\nulldelimiterspace 
        #1 
        \vphantom{|} 
        \right|_{#2}^{#3} 
}}

\usepackage{acro}

\DeclareAcronym{DG}
{
    short = DG ,
    long = discontinuous Galerkin
}

\DeclareAcronym{ACA}
{
    short = ACA ,
    long = adaptive cross approximation
}

\DeclareAcronym{EFIE}
{
    short =  EFIE ,
    long = electric field integral equation
}

\DeclareAcronym{MFIE}
{
    short =  MFIE ,
    long = magnetic field integral equation
}

\DeclareAcronym{CFIE}
{
    short =  CFIE ,
    long = combined field integral equation
}

\DeclareAcronym{MUIE}
{
    short =  MUIE ,
    long = Müller integral equation
}

\DeclareAcronym{PMCHWT}
{
    short =  PMCHWT ,
    long = Poggio-Miller-Chang-Harrington-Wu-Tsai integral equation
}

\DeclareAcronym{SPD}
{
    short =  SPD ,
    long = {symmetric, positive definite}
}

\DeclareAcronym{SPSD}
{
    short =  SPD ,
    long = {symmetric, positive semi-definite}
}

\DeclareAcronym{PEC}
{
    short =  PEC ,
    long = perfectly electrically conducting
}

\DeclareAcronym{RWG}
{
    short = RWG ,
    long = Rao-Wilton-Glisson
} 

\DeclareAcronym{BC}
{
    short = BC ,
    long = Buffa-Christiansen
}

\DeclareAcronym{SVD}
{
    short = SVD ,
    long = singular value decomposition
}

\DeclareAcronym{CG}
{
    short = CG ,
    long = conjugate gradient
} 

\DeclareAcronym{PCG}
{
    short = PCG ,
    long = preconditioned conjugate gradient
} 

\DeclareAcronym{CGS}
{
    short = CGS ,
    long = conjugate gradient squared
}

\DeclareAcronym{CMP}
{
    short = CMP ,
    long = Calderón multiplicative preconditioner
} 

\DeclareAcronym{RFCMP}
{
    short = RF-CMP ,
    long = refinement-free Calderón multiplicative preconditioner
} 

\DeclareAcronym{HPD}
{
    short = HPD ,
    long = {Hermitian, positive definite}
} 

\DeclareAcronym{RHS}
{
    short = RHS ,
    long = right-hand side
}

\DeclareAcronym{LSE}
{
    short = LSE ,
    long = linear system of equations
} 

\newcolumntype {n}{c}
\newcolumntype {N}{>{\small}c}
\newcolumntype {L}{>{\small}l}
\newcolumntype {F}{>{\footnotesize}c}
\newcolumntype {v}[1]{>{\raggedright \hspace {0pt}} p {#1}}
\newcolumntype {V}[1]{>{\small \raggedright \hspace {0pt}} p {#1}}
\newcolumntype{d}[1]{>{\DC@{.}{.}{#1}}c<{\DC@end}}

\newcolumntype{R}[1]{%
    >{\begin{turn}{90}\begin{minipage}{#1}\small\raggedright\hspace{0pt}}l%
            <{\end{minipage}\end{turn}}%
}

\NewDocumentCommand{\TA}{o}{
    \IfNoValueTF {#1} {%
        \vecop T_{\kern-2pt\mr{A}}
    }
    {
        \vecop T_{\kern-2pt\mr{A},#1}
    }
}

\NewDocumentCommand{\TPhi}{o}{
    \IfNoValueTF {#1} {%
        \vecop T_{\kern-2pt\Phiup}
    }
    {
        \vecop T_{\kern-2pt\Phiup,#1}
    }
}

\NewDocumentCommand{\matTA}{o}{
    \IfNoValueTF {#1} {%
        \mat T_\mr{A}   
        }
    {
        \mat T_{\mr{A},#1}
    }
}

\NewDocumentCommand{\matTPhi}{o}{
    \IfNoValueTF {#1} {%
        \mat T_\Phiup   
        }
    {
        \mat T_{\Phiup,#1}
    }
}

\NewDocumentCommand{\MSL}{o}{
    \IfNoValueTF {#1} {%
        \veg \Psi_\mr{SL}
        }
    {
        \veg \Psi_{\mr{SL},#1}
    }
}

\NewDocumentCommand{\MDL}{o}{
    \IfNoValueTF {#1} {%
        \veg \Psi_\mr{DL}
        }
    {
        \veg \Psi_{\mr{DL},#1}
    }
}

\NewDocumentCommand{\PA}{o}{
    \IfNoValueTF {#1} {%
        \veg \Psi_\mr{A}
        }
    {
        \veg \Psi_{\mr{A},#1}
    }
}

\NewDocumentCommand{\PPhi}{o}{
    \IfNoValueTF {#1} {%
        \veg \Psi_{\Phiup}
        }
    {
        \veg \Psi_{\Phiup,#1}
    }
}

\usepackage{nicefrac}
\usepackage{lipsum} 
\usepackage{shellesc}
\usepackage{multirow}

\newcommand{\dyad}[1]{\boldsymbol{\bar{#1}}}
\renewcommand{\vec}[1]{\boldsymbol{#1}}

\graphicspath{{./figures/}}  

\usepackage{ifluatex}
\ifluatex

\usepackage{luacode}
\usepackage[strings]{underscore}  

\input{ExternalizingFigures.lua}

\directlua{RecompileAlteredFigures("./figures", "LuaLatex", "MyFigureStyle.sty", 0)}
\fi

\ifCLASSOPTIONcompsoc
 \usepackage[caption=false,font=normalsize,labelfont=sf,textfont=sf]{subfig}
\else
 \usepackage[caption=false,font=footnotesize]{subfig}
\fi

 \usepackage{atbegshi}
\makeatletter
\def\ps@IEEEtitlepagestyle{%
	\def\@oddfoot{\mycopyrightnotice}%
	\def\@oddhead{\hbox{}\@IEEEheaderstyle\leftmark\hfil\thepage}\relax
	\def\@evenhead{\@IEEEheaderstyle\thepage\hfil\leftmark\hbox{}}\relax
	\def\@evenfoot{}%
}
\AtBeginShipout{\AtBeginShipoutUpperLeft{%
		\put(\dimexpr\paperwidth-1cm\relax,-.55cm){\makebox[0pt][r]{
				\begin{minipage}{\textwidth}
					\centering \scriptsize
					This article has been accepted for publication in IEEE. This is the author's version which has not been fully edited and content may change prior to final publication. Citation information: DOI 10.23919/EuCAP63536.2025.10999865
				\end{minipage}
	}}}%
}
\def\mycopyrightnotice{%
	\begin{minipage}{\textwidth}
		\centering \scriptsize
		Copyright~\copyright~2025 IEEE. Personal use of this material is permitted. Permission from IEEE must be obtained for all other uses, in any current or future media, including\\reprinting/republishing this material for advertising or promotional purposes, creating new collective works, for resale or redistribution to servers or lists, or reuse of any copyrighted component of this work in other works by sending a request to pubs-permissions@ieee.org.
	\end{minipage}
}

\begin{document}
%
\title{Near-Field Focusing Operators for Planar Multi-Static Microwave Imaging Using Back-Projection in the Spatial Domain}

\author{\IEEEauthorblockN{
Matthias M. Saurer\IEEEauthorrefmark{1},   
Marius Brinkmann\IEEEauthorrefmark{2}, 
Han Na\IEEEauthorrefmark{1}, 
Quanfeng Wang\IEEEauthorrefmark{1},   
and  
Thomas F. Eibert\IEEEauthorrefmark{1}   
}                                     
\IEEEauthorblockA{\IEEEauthorrefmark{1}
Department of Electrical Engineering, School of Computation, Information and Technology, \\ Technical University of Munich, Munich, Germany (hft@ei.tum.de)\\}
\IEEEauthorblockA{\IEEEauthorrefmark{2}
Rohde \& Schwarz GmbH \& Co. KG, Munich, Germany (marius.brinkmann@rohde-schwarz.com)}
}



\maketitle

\begin{abstract}
	Based on a plane-wave expansion of the observation data in quasi-planar multi-static scattering scenarios, an improved formalism for image creation utilizing back-projection in the spatial domain is derived. The underlying integral expressions for different focusing operators are derived analytically leading to magnitude correction factors, which are mostly relevant for reconstructing microwave images when the distance from the scattering object to the aperture plane is small. It is shown that the derived imaging procedure is superior to the traditional back-projection only compensating the phase delay of the measurement signals and validate our findings based on simulated as well as measured data. Since the derived focusing operators correspond to a low-pass filtering of the spatial images, the resulting modified multi-static back-projection algorithms effectively suppress imaging artifacts as well. 
\end{abstract}

\vskip0.5\baselineskip

%

\section{Introduction}
The back-projection algorithm (BPA) has been proven as a very reliable and stable method for creating holographic microwave images~\cite{Saurer.Sep.2022}. Conceptually the BPA is quite intuitive and is applied to measurements for a set of discrete observation configurations. In the BPA, the measured
field values are multiplied with so called focusing operators, which account for the wave propagation between the scattering object and the transmitters (Tx) and receivers (Rx)~\cite{Osipov.Feb.2013}, and the spatial image is obtained by coherently superimposing the back-projected field values. If the field samples are taken in the far field of the scattering object, each Tx and Rx pair can be directly linked to the propagation of a plane wave and the back-projection simply reduces to a phase correction compensating the phase delay of the different plane waves. This plane-wave concept can also be applied to near-field (NF) scattering scenarios and is simply based on an expansion of the observation data into plane waves utilizing Fourier integrals. Assuming the validity of the first-order Born approximation and a closed observation surface around the scattering object, this approach allows for a very efficient solution of the inverse scattering problem~\cite{Devaney.2012}, since the plane-wave expansion leads to a diagonalization of the forward scattering operator. Compared to the spatial back-projection, this plane-wave based back-projection approach requires an additional transition from the spatial domain towards the spectral domain and vice versa and is for instance described in \cite{Saurer.Mar.2022}. Utilizing hierarchical methods these transitions can be computed very effectively also for irregularly distributed measurement locations. However, for sparse antenna arrays, often aliasing artifacts are introduced, which can significantly deteriorate the overall image quality. This rises the interesting question, whether spatial back-projection can also be applied to NF scattering scenarios and which correction factors should be applied with respect to the magnitude decay of the measurement signals. Since spatial and spectral back-projection are, as already mentioned, simply related by Fourier integrals, we can derive an explicit expression for the required focusing operators and analytically solve the underlying integral expression with consideration of low-pass filtering functions of various orders, which can help to improve the image qualtiy. Based on simulated and measured data it is demonstrated that utilizing the derived focusing operators lead to a considerable improvement of the reconstructed microwave images in NF scattering scenarios.

\section{Theoretical derivation}
For a given position of the Tx, denoted by $\vec{r}_{\mathrm{T}}$, and the Rx, denoted by $\vec{r}_{\mathrm{R}}$, both located in the plane $z=z_m>0$,
measurements described by the scattering parameter $T\left(\vec{r}_{\mathrm{R}},\vec{r}_{\mathrm{T}}\right)$ are performed. Utilizing the planar plane-wave representation of the Weyl-identity to equivalently represent the kernel of the Green's function~\cite{Saurer.Mar.2022}, this scattering parameter can be expanded into a superposition of incident and scattered plane waves with wave vectors \mbox{$\vec{k}^{\mathrm{i}}=\left[k_x^{\mathrm{i}},k_y^{\mathrm{i}},-k_z^{\mathrm{i}}\right]^{\mathrm{T}}$} and  \mbox{$\vec{k}^{\mathrm{s}}=\left[k_x^{\mathrm{s}},k_y^{\mathrm{s}},k_z^{\mathrm{s}}\right]^{\mathrm{T}}$} according to
\begin{multline}\label{gen_expr}	
	T\left(\vec{r}_{\mathrm{R}},\vec{r}_{\mathrm{T}}\right)=\iint\limits_{-\infty}^{~~+\infty}\iint\limits_{-\infty}^{~~+\infty}
	\tilde{\vec{W}}_{\mathrm{R}}\left(-\vec{k}^{\mathrm{s}}\right)\cdot
	\frac{	\tilde{\bar{\vec{s}}}_{\mathrm{B}}\left(\vec{k}^{\mathrm{s}}-\vec{k}^{\mathrm{i}}\right)}{k^2k_z^{\mathrm{s}}k_z^{\mathrm{i}}}
\cdot\tilde{\vec{W}}_{\mathrm{T}}\left(\vec{k}^{\mathrm{i}}\right)\\
\e^{-\jm\vec{k}^{\mathrm{s}}\cdot\vec{r}_{\mathrm{R}}}
\e^{+\jm \vec{k}^{\mathrm{i}}\cdot\vec{r}_{\mathrm{T}}}	\,\mathrm{d}k_x^{\mathrm{s}}\,\mathrm{d}k_y^{\mathrm{s}}\,\mathrm{d}k_x^{\mathrm{i}}\,\mathrm{d}k_y^{\mathrm{i}}\,,
\end{multline}
where
\begin{equation}\label{k_case}
	k_z=\begin{cases}
		\sqrt{k^2-k_x^2-k_y^2} & \text{for $ k^2>k_x^2+k_y^2 $}\\
		-\jm\sqrt{k_x^2+k_y^2-k^2} & \text{for $k^2<k_x^2+k_y^2$}
	\end{cases}\,
\end{equation}
denotes the dispersion relation of the wave number \mbox{$k=\abs{\vec{k}}=\sqrt{k_x^2+k_y^2+k_z^2}$}, which segments the planar plane-wave spectrum into a visible region with $k^2>k_x^2+k_y^2$ and an evanescent region for $k^2<k_x^2+k_y^2$. In (\ref{gen_expr}), $\tilde{\vec{W}}_{\mathrm{R}}(-\vec{k}^{\mathrm{s}})$ and $\tilde{\vec{W}}_{\mathrm{T}}(\vec{k}^{\mathrm{i}})$ are the plane-wave representations
of the Rx and Tx antenna, respectively, and are obtained by their corresponding electric current densities via
\begin{equation}
	\tilde{\vec{W}}_{\mathrm{T/R}}\left(\vec{k}\right)=\frac{-Zk^2 }{8\uppi^2}\left(\mathbf{\bar{I}}-\hat{\vec{k}}\hat{\vec{k}}\right)\cdot\iiint\limits_{V_{\mathrm{T/R}}}\vec{J}_{\mathrm{T/R}}(\vec r')\e^{\,\jm \vec{k}\cdot\vec{r}'}\,\mathrm{d}^3 \vec{r}'\,,\label{wtr}
\end{equation}
where $Z=\sqrt{\mu/\varepsilon}$ denotes the wave impedance of the underlying homogeneous space with permeability $\mu$ and permittivity $\varepsilon$ and $\hat{\vec{k}}$ represents the wave vector normalized to its unit length. In (\ref{gen_expr}), $\tilde{\bar{\vec{s}}}_{\mathrm{B}}(\vec{k}^{\mathrm{s}}-\vec{k}^{\mathrm{i}})$ denotes the scattering dyad in the spectral domain under the first-order Born approximation. The spatial single-frequency image $\dyad{s}_{\mathrm{B}}(\vec{r}')$ is computed by collecting the available plane-wave contributions according to
\begin{multline}\label{img_gen}
	\dyad{s}_{\mathrm{B}}\left(\vec{r}'\right)= {\left(\frac{1}{2\uppi}\right)}^4
	\iint\limits_{-\infty}^{~+\infty}\iint\limits_{-\infty}^{~+\infty}H(\vec{k}^{\mathrm{s}})H(-\vec{k}^{\mathrm{i}})\frac{\tilde{\bar{\vec{s}}}_{\mathrm{B}}(\vec{k}^{\mathrm{s}}-\vec{k}^{\mathrm{i}})}{k^2k_z^{\mathrm{s}}k_z^{\mathrm{i}}}\\\e^{-\jm
		\left(\vec{k}^{\mathrm{s}}-\vec{k}^{\mathrm{i}}\right)\cdot\vec{r}'}\,\mathrm{d}^2\vec{k}^{\mathrm{s}}\,\mathrm{d}^2\vec{k}^{\mathrm{i}}\,,
\end{multline}
where $H(\vec{k})$ is a filter function introduced to truncate the doubly infinite integrals and mitigate corresponding artifacts~\cite{Devaney.2012}. Interpreting (\ref{gen_expr}) as an inverse 4-D Fourier transform, the corresponding spectral representation of the transfer operator is identified as 
\begin{equation}\label{spec_eq}
	{\left(\frac{1}{2\uppi}\right)}^4\tilde{T}(\vec{k}^{\mathrm{s}},-\vec{k}^{\mathrm{i}})=\tilde{\vec{W}}_{\mathrm{R}}(-\vec{k}^{\mathrm{s}})\cdot\frac{\tilde{\bar{\vec{s}}}_{\mathrm{B}}(\vec{k}^{\mathrm{s}}-\vec{k}^{\mathrm{i}})}{k^2k_z^{\mathrm{s}}k_z^{\mathrm{i}}}\cdot\tilde{\vec{W}}_{\mathrm{T}}(\vec{k}^{\mathrm{i}})\,,
\end{equation}
where the left hand side of this equation contains the spectral representation of the measurement signals, which is computed according to
\begin{equation}
	\label{spec_comp}
	\tilde{T}(\vec{k}^{\mathrm{s}},-\vec{k}^{\mathrm{i}})=\iint\limits_{-\infty}^{~+\infty}\iint\limits_{-\infty}^{~+\infty}T\left(\vec{r}_{\mathrm{R}},\vec{r}_{\mathrm{T}}\right)\e^{\,\jm
		\vec{k}^{\mathrm{s}}\cdot\vec{r}_{\mathrm{R}}}
	\e^{-\jm \vec{k}^{\mathrm{i}}\cdot\vec{r}_{\mathrm{T}}}\,\mathrm{d}^2\vec{r}_{\mathrm{T}}\,\mathrm{d}^2\vec{r}_{\mathrm{R}}\,.
\end{equation}
The right hand side of (\ref{spec_eq}) is reduced to a scalar scattering coefficient, if one dominant polarization and isotropic antennas are assumed. Combining (\ref{img_gen}), (\ref{spec_eq}), and (\ref{spec_comp}), and ignoring some not so relevant constants results then into
\begin{multline}\label{foc_def}
	s_{\mathrm{B}}\left(\vec{r}'\right)=\iint\limits_{-\infty}^{~+\infty}\iint\limits_{-\infty}^{~+\infty}T\left(\vec{r}_{\mathrm{R}},\vec{r}_{\mathrm{T}}\right)
	\underbrace{\iint\limits_{-\infty}^{~+\infty}H(\vec{k}^{\mathrm{s}})\e^{-\jm
			\vec{k}^{\mathrm{s}}\cdot\left(\vec{r}'-\vec{r}_{\mathrm{R}}\right)}\,\mathrm{d}^2\vec{k}^{\mathrm{s}}}_{F\left(\vec{r}'-\vec{r}_{\mathrm{R}},k\right)}\\
	\underbrace{\iint\limits_{-\infty}^{~+\infty}H(-\vec{k}^{\mathrm{i}})\e^{+\jm \vec{k}^{\mathrm{i}}\cdot\left(\vec{r}'-\vec{r}_{\mathrm{T}}\right)}\,\mathrm{d}^2\vec{k}^{\mathrm{i}}}_{F\left(\vec{r}'-\vec{r}_{\mathrm{T}},k\right)}\,\mathrm{d}^2\vec{r}_{\mathrm{R}}\,\mathrm{d}^2\vec{r}_{\mathrm{T}}\,.
\end{multline}
Obviously the integral expression
\begin{equation}\label{foc_def2}
	F\left(\vec{r}'-\vec{r},k\right)=\iint\limits_{-\infty}^{~+\infty}H(\vec{k})\e^{-\jm \vec{k}\cdot\left(\vec{r}'-\vec{r}\right)}\,\mathrm{d}^2\vec{k}
\end{equation}	
defines the focusing operators for the transmit and receive array with respect to a specific pixel position $\vec{r}'$, where $H(-\vec{k})=H(\vec{k})$ is assumed.
With these definitions, the modified multi-static back-projection algorithm for a single-frequency image according 
to
\begin{multline}\label{bpa_mod_eq}
	s_{\mathrm{B}}\left(\vec{r}'\right)=\iint\limits_{-\infty}^{~+\infty}\iint\limits_{-\infty}^{~+\infty}T\left(\vec{r}_{\mathrm{R}},\vec{r}_{\mathrm{T}}\right)\\
	F\left(\vec{r}'-\vec{r}_{\mathrm{R}},k\right)F\left(\vec{r}'-\vec{r}_{\mathrm{T}},k\right)\,\,\mathrm{d}^2\vec{r}_{\mathrm{R}}
	\,\mathrm{d}^2\vec{r}_{\mathrm{T}}\,
\end{multline}
can be formulated. The multi-frequency image is simply obtained by coherently superimposing all the individual single-frequency images followed by a normalization to the maximum intensity value. Employing the focusing operator, which was for a monostatic setup first introduced in~\cite{Saurer.Jan.2025}, allows to perform the image creation directly in the spatial domain and, hence, completely avoids the transition into the $k$-space. For the filter functions of the form $H_n(\vec{k})=k_z^n$ with $n\in\mathbb{N}_0$ a closed form expression for the integral in (\ref{foc_def2}) is conveniently obtained by employing the Weyl-identity in its planar plane-wave form given by
\begin{equation}\label{weyl_id}
	\frac{\e^{-\mathrm{j}k\abs{\vec{R}}}}{\abs{\vec{R}}}=\iint\limits_{-\infty}^{\infty}\frac{\e^{-\mathrm{j}\left(k_xR_x+k_yR_y\right)}}{2\uppi\mathrm{j}k_z}\e^{-\mathrm{j}k_z\abs{R_z}}\,\mathrm{d}^2\vec{k}\,,
\end{equation}
where  $\vec{R}=\vec{r}'-\vec{r}=R_x\hat{\vec{x}}+R_y\hat{\vec{y}}+R_z\hat{\vec{z}}$ and  $R=\sqrt{R_x^2+R_y^2+R_z^2}$. Employing the derivative theorem of the Fourier integrals it can be concluded
\begin{equation}\label{comp_foc_mult}
	F_{n}\left(\vec{R},k\right)=\begin{cases}\frac{-2\uppi}{{\left(-\mathrm{j}\right)^n}}\frac{\partial^{n+1}}{ {\left(\partial R_z\right)}^{n+1}}\frac{\e^{-\mathrm{j}kR}}{R}\,,R_z>0\\
		\frac{+2\uppi}{{\left(-\mathrm{j}\right)^n}}\frac{\partial^{n+1}}{ {\left(\partial R_z\right)}^{n+1}}\frac{\e^{+\mathrm{j}kR}}{R}\,,R_z<0
	\end{cases}
\end{equation}
which leads for the cases $n=0,1,2$ and the assumption of $R_z<0$ as considered throughout this work to the explicit computation of the three focusing operators
\begin{equation}\label{foc_op_n_0}
	F_{0}\left(\vec{R},k\right) =2\uppi \left(\frac{\mathrm{j}kR_z}{R^2}-\frac{R_z}{R^3}\right)\e^{\,\mathrm{j}kR}\,,~~~~~~~~~~~~~~~~~~~~~
\end{equation}
\begin{multline}
	\label{foc_op_n_1}F_{1}\left(\vec{R},k\right)=2\uppi\jm	\left[\frac{\mathrm{j}^2k^2R_z^2}{R^3}+\left(R_x^2+R_y^2-2R_z^2\right)\right.\phantom{~~~}\\\left.\left(\frac{\mathrm{j}k}{R^4}-\frac{1}{R^5}\right)\right]\e^{\,\mathrm{j}kR}\,,
\end{multline}
\begin{multline}\label{foc_op_n_2}
	F_{2}\left(\vec{R},k\right)=-2\uppi\left[\frac{\mathrm{j}^3{k}^3R_z^3}{R^4}+R_z\left(\frac{3\mathrm{j}^2{k}^2(R_x^2+R_y^2-R_z^2)}{R^5}+\right.\right.\\
	\left.\left.(3R_x^2+3R_y^2-2R_z^2)\left(-\frac{3\mathrm{j}k}{R^6}+\frac{3}{R^7}\right)\right)\right]\e^{\,\mathrm{j}kR}\,.
\end{multline}

\section{Simulation and measurement results}
\subsection{Ideal Point Scatterers}
\begin{figure}
	\centering
	\includegraphics[scale=0.83]{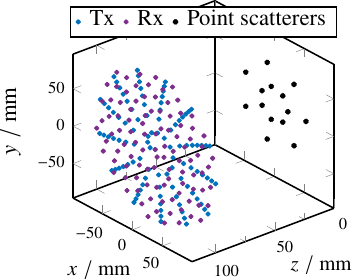}\hfill
	\caption{Visualization of the first imaging scenario containing 14 isotropic point scatterers. The simulation data in this case was synthetically generated using Hertzian dipoles for the Tx and Rx.}
	\label{vis_Pt_scat}
\end{figure}

\begin{figure}
	\centering
	\subfloat[]{\includegraphics[scale=0.74]{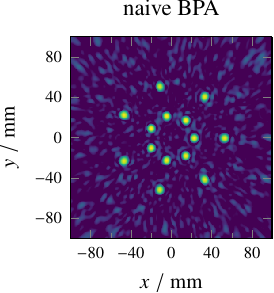}}\hfill
	\subfloat[]{\includegraphics[scale=0.74]{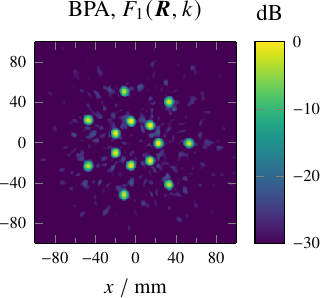}}
	\caption{Reconstruction results for the first simulation including 14 discrete point scatterers and ideal Hertzian dipole probes for the transmitting and receiving antennas. (a) Standard BPA without magnitude correction, (b) improved BPA.} 
	\label{Pt_Scat_results_v1}
\end{figure}

In order to verify the validity of the proposed magnitude correction terms, a simulation based on discrete point scatterers as shown in Fig.~\ref{vis_Pt_scat} was performed. 
The utilized Tx and Rx in this case underlie a spiral sampling grid and are, hence, irregularly distributed on an aperture circle with maximum radius of 10\,cm. A total of 14 ideal (isotropic) point scatterers are placed 10\,cm away from the antenna positions in two circles of radii 2.25 cm and 5.25 cm with seven scatterers for each of the two circles. To better demonstrate the improved focusing capabilities of the derived focusing operators, this simulation was only conducted at a single  frequency, namely 40\,GHz. It can be clearly seen that utilizing the BPA with the magnitude correction factor $\abs{F_1\left({\vec{R},k}\right)}$ leads to a significant improvement in the overall image quality. As can be seen in Fig.~\ref{Pt_Scat_results_v1}(b), the level of artifacts is strongly decreased compared to the result from the conventional back-propagation method as seen in Fig.~\ref{Pt_Scat_results_v1}(a) without employing any magnitude correction factors. 
\subsection{Full-Wave Simulation in CST MWS}
\begin{figure}
	\centering
	\includegraphics[scale=0.83]{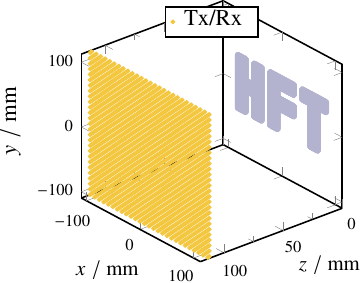}\hfill
	\caption{Scattering scenario simulated in CST MWS utilizing planar measurements and the perfectly electrically conducting HFT logo.}
	\label{vis_HFT}
\end{figure}

The next simulation was carried out in CST MWS utilizing the integral equation solver at a single frequency of 40\,GHz employing a multi-static setup of 6400 Tx and 6400 Rx spanning an aperture size of 22\,cm by 22\,cm with regular sampling. This aperture is located in a plane with a distance of 11\,cm to a perfectly electrically conducting HFT logo as can be seen in Fig.~\ref{vis_HFT}.
Accurately evaluating the integral in (\ref{bpa_mod_eq}) while utilizing the correct quadrature weights (which are trivial in this case) and the focusing operator of lowest order given by $F_0\left(\vec{R},k\right)$ in (\ref{foc_op_n_0}) is mathematical equivalent to (\ref{img_gen}) without any spectral filtering and, hence, closely related to the MIMO-$\omega$-$k$ presented in~\cite{Wang.Jul.2020}.
\begin{figure}
	\centering
	\subfloat[]{\includegraphics[scale=0.76]{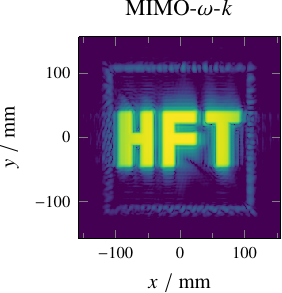}}\hfill
	\subfloat[]{\includegraphics[scale=0.76]{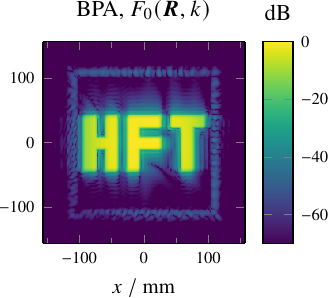}}\\
	\subfloat[]{\includegraphics[scale=0.76]{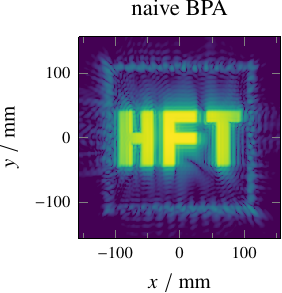}}\hfill
	\subfloat[]{\includegraphics[scale=0.76]{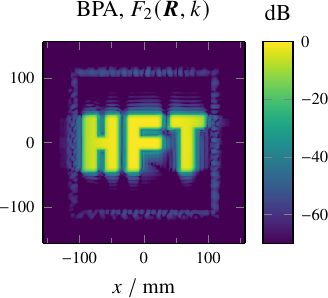}}
	\caption{Reconstruction results for the HFT logo. (a) Spatial image of the unfiltered MIMO-$\omega$-$k$-algorithm utilizing the code provided with~\cite{Wang.Jul.2020}, (b),(c),(d) back-projection algorithm.} 
	\label{CST_reks}
\end{figure}
As depicted in Fig.~\ref{CST_reks}(a) and (b), respectively, the spatial images computed by the MIMO-$\omega$-$k$ imaging algorithm (while ignoring the terms $k_{y_{\mathrm{t}}}$ and $k_{y_{\mathrm{r}}}$ of (10) in~\cite{Wang.Jul.2020}) and the image obtained by the modified BPA with the focusing operator $F_0\left(\vec{R},k\right)$ both reconstructed at $z=0$ show an excellent agreement as well as a very good focusing of the HFT logo. Compared to this, the standard BPA compensating only the phase of the measurement data leads to strong artifacts in the entire imaging domain as shown in
Fig.~\ref{CST_reks}(c). These artifacts are again greatly suppressed when employing the second-order focusing operator $F_2\left(\vec{R},k\right)$ according to (\ref{foc_op_n_2}) as shown in Fig.~\ref{CST_reks}(d). Due to the quadratic truncation of the aperture for the transmit and receive array, artifacts close to all four edges of the observation domain are introduced in all reconstructed microwave images. 

\subsection{Ray-Tracing Simulation of Metallic Plate}
\begin{figure}
	\centering
	\subfloat[]{\includegraphics[scale=0.75]{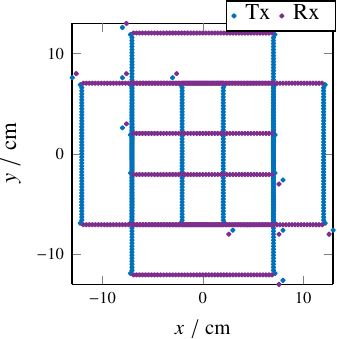}}\hspace{1cm}
	\subfloat[]{\includegraphics[scale = 0.17]{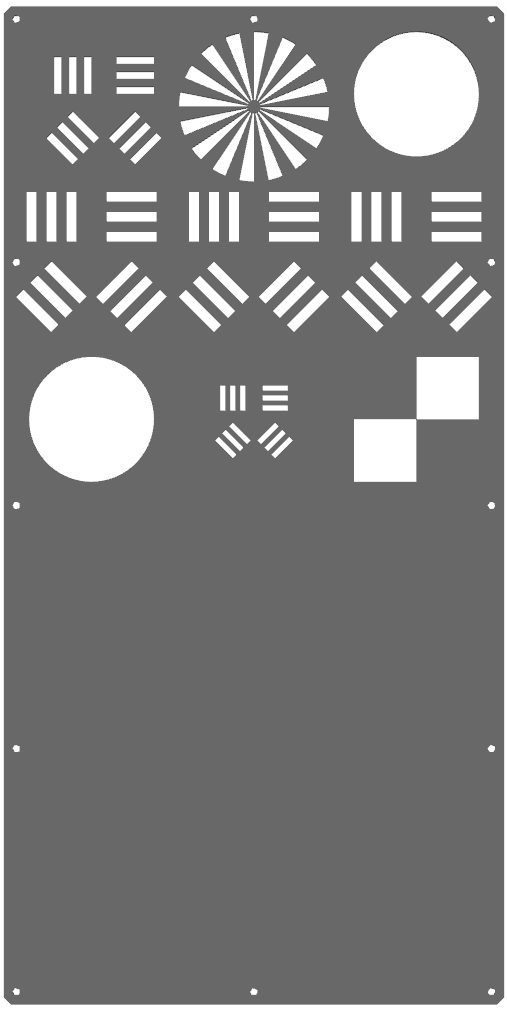}}
	\caption{Visualization of the utilized Tx and Rx positions in the GO based ray-tracing simulation (a) as well as of the metallic plate containing several evaluation patterns (b).}
	\label{ISAR_pos}
\end{figure}

\begin{figure}
	\centering
	\subfloat[]{\includegraphics[scale=0.75]{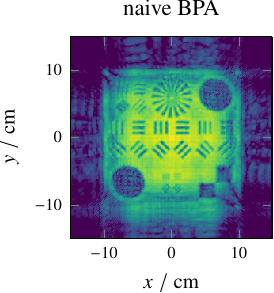}}\hfill
	\subfloat[]{\includegraphics[scale=0.75]{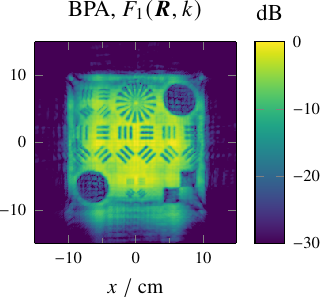}}\\
	\subfloat[]{\includegraphics[scale=0.75]{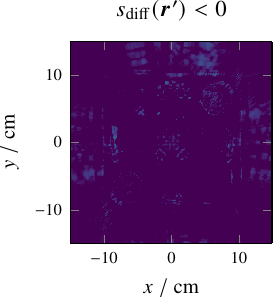}}\hfill
	\subfloat[]{\includegraphics[scale=0.75]{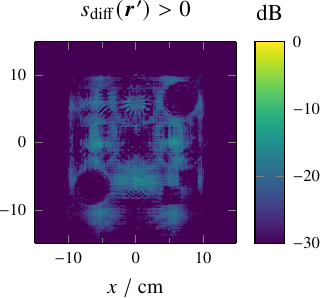}}
	\caption{Maximum intensity projection of the volumetric reconstruction results along the $z$-dimension. (a) and (b) show the spatial images for the standard BPA as well as the modified  BPA utilizing the focusing operator $F_1\left(\vec{R},k\right)$. (c) and (d) highlight the positive and negative variations in the magnitude behavior of the scattering object in the difference image.} 
	\label{Plate_RuS_results}
\end{figure}

The next simulation is based on a metallic plate containing several evaluation patterns like rectangular and circular cutouts of different size and shape. The underlying array configuration corresponds to the Rohde \& Schwarz QAR50~\cite{Brinkmann.Sep.2023}, where, however, multiple simulation results with different relative positions between the antenna array and the metallic plate according to a synthetic aperture radar (SAR) approach were combined in order to effectively increase the achievable imaging resolution. Hence, a total of 480 Tx and 480 Rx as well as 128 linearly distributed frequency samples from 71\,GHz to 81\,GHz were employed in this simulation, which was conducted utilizing a geometrical optics (GO) based ray-tracer~\cite{Schnattinger.Sep.2023}. A schematic visualization of the antenna positions as well as of the metallic plate serving as the scattering object in this case is given in Fig.~\ref{ISAR_pos}(a) and (b).
The corresponding reconstruction results for this case are given in Fig.~\ref{Plate_RuS_results}(a) for the standard BPA and in Fig.~\ref{Plate_RuS_results}(b) for the modified BPA including the magnitude correction factor for the focusing operator of first order $F_1\left(\vec{R},k\right)$ as given in (\ref{foc_op_n_1}). To better visualize the discrepancies in the two images,  the difference image $s_{\mathrm{diff}}(\vec{r}')$ according to $s_{\mathrm{diff}}(\vec{r}') = \abs{s_{\mathrm{BPA, mod}}(\vec{r}')} -\,\abs{s_{\mathrm{BPA}}(\vec{r}')}$ is computed, where the nomenclature of the subscripts should be self-explanatory. Based on Fig.~\ref{Plate_RuS_results}(c) it can be stated that the modified BPA effectively suppresses the overall level of artifacts especially outside of the metallic plate. Meanwhile the difference image in Fig.~\ref{Plate_RuS_results}(d) shows that the inner part of the target has larger magnitudes for the modified BPA and is, therefore, better focused. Overall, this greatly demonstrates that the utilization of the derived NF focusing operators including magnitude correction terms lead to better contrast in the reconstructed images and strongly mitigates artifacts in the entire imaging domain.
\subsection{Imaging of a Human Person}
\begin{figure}
	\centering
	\includegraphics[scale=0.72]{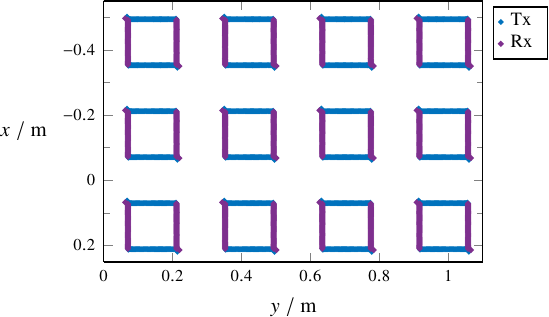}
	\caption{Tx and Rx placement for the Rohde \& Schwarz QAR as utilized in the conducted measurement of imaging a human person.}
	\label{QPS_array}
\end{figure}

\begin{figure}
	\centering
	\subfloat[]{\includegraphics[scale=0.76]{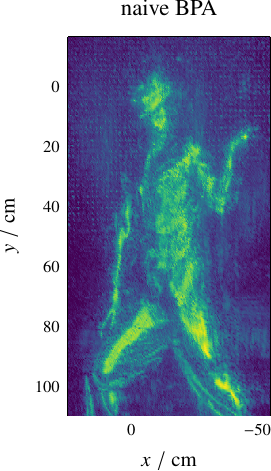}}\hfill
	\subfloat[]{\includegraphics[scale=0.76]{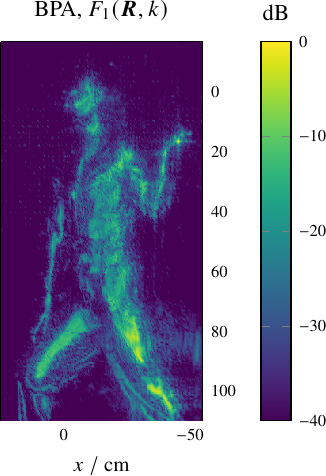}}
	\caption{Maximum intensity projection along the $z$-axis of the reconstructed spatial image of a real person.} 
	\label{Lee_xy}
\end{figure}

\begin{figure}[!htbp]
	\centering
	\subfloat[]{\includegraphics[scale=0.76]{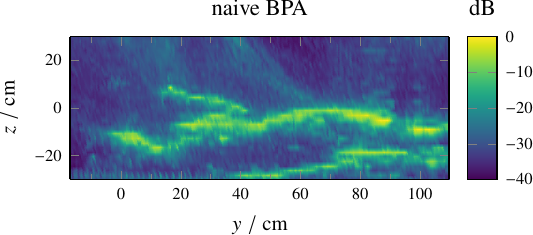}}\\
	\subfloat[]{\includegraphics[scale=0.76]{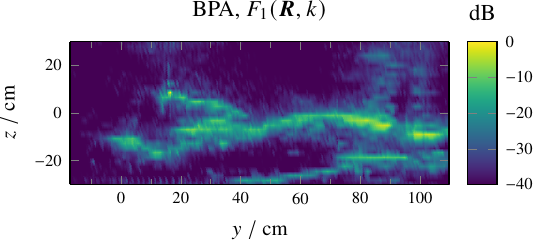}}
	\caption{Maximum intensity projection along the $x$-axis of the reconstructed spatial image of a real person.} 
	\label{Lee_yz}
\end{figure}

Lastly,  the performance of the BPA with the proposed NF correction factors with respect to real measurement data when imaging a human person is investigated utilizing the Rohde \& Schwarz QAR quality automotive radome tester~\cite{.b}. This device consists of 4 by 3 clusters each containing 96 Tx and 96 Rx, which are depicted in Fig.~\ref{QPS_array} and the 128 frequency samples were again linearly distributed in the frequency range from 71 GHz to 81 GHz.
The spatial images were reconstructed in a volume of size 80\,cm by 160\,cm by 60\,cm with a resolution of 256 by 256 by 32 pixels utilizing a CUDA based implementation of the different considered versions of the BPA. Employing an NVIDIA RTX A6000 GPU, the time for the image creation for the standard BPA was 5\,h, while the total running time for the modified BPA with the focusing operator $F_1\left(\vec{R},k\right)$ was approximately 10\,h. In Fig.~\ref{Lee_xy} and Fig.~\ref{Lee_yz}, the maximum intensity projections along the $z$- and $x$- axis for the two methods are shown. As observed in previous simulations the utilization of focusing operators as derived in this work leads to a strong reduction of the artifacts in the entire imaging domain as can be seen in Fig.~\ref{Lee_xy}(b) and Fig.~\ref{Lee_yz}(b), respectively. This finding is confirmed by evaluating the image entropy, which yields a value of -\,9.80 for the improved imaging algorithm including magnitude correction and low-pass filtering and -10.84 for the naive BPA. However, it also seems that the intensity values on most parts on the surface of the human person are for the proposed method also a little bit weaker compared to the standard BPA. However, this is due to the configuration of the antenna array as given in Fig.~\ref{QPS_array} and the strong directivity of the focusing operator. Since the occupancy  of the Tx and Rx with respect to the size of the imaging domain is relatively small, the refocusing of the observation data is also limited and has less selectivity for pixel positions, which projection into the aperture plane have larger distances to any of the Tx/Rx pairs. A possible workaround could be the combination of several measurements with slightly different relative position and utilizing the SAR approach similarly as for the previously mentioned simulation of a metallic plate.
\section{Conclusion}
Magnitude correction factors for near-field multi-static microwave imaging utilizing back-projection in the spatial domain were presented. Based on a plane-wave representation of the inverse scattering problem under the first-order Born approximation, different focusing operators, which can directly be applied to the observation data in spatial domain, were derived. It was demonstrated that utilizing these focusing operators, which are equivalent to a low-pass filtering of the spatial images, can considerably increase the dynamic range of the reconstruction while simultaneously leading to a noticeable decrease of the level of artifacts in short-range imaging applications. For validation of our theoretical considerations, different types of simulations as well as measured data involving a human person were utilized.  


\section*{Acknowledgment}
Funded by the European Union. Views and opinions expressed are however those of the author(s) only and do not necessarily reflect those of the European Union or European Innovation Council and SMEs Executive Agency (EISMEA). Neither the European Union nor the granting authority can be held responsible for them. Grant Agreement No:~101099491.

\bibliographystyle{IEEEtran}
\bibliography{Literature1}

\end{document}